\def\arXiv{}
\ifx\draft\undefined
	\let\sidebysidefigures=\undefined
	\documentclass[journal, twocolumn, twoside]{IEEEtran}
\else
	\documentclass[journal, draftclsnofoot, 12pt, onecolumn, twoside]{IEEEtran}
\fi
\pdfoutput=1

\usepackage{amsmath}
\usepackage{amssymb}
\usepackage{amsthm}
\usepackage{bbm}\let\mathbb\mathbbmss
\usepackage{graphicx}\graphicspath{{./figs/}}
\usepackage[caption=false]{subfig}
\usepackage{cite}
\usepackage{booktabs}
\usepackage{microtype}
\usepackage[hyphens]{url}
\usepackage{multirow}

\ifx\arXiv\undefined\else

\usepackage{color}
\definecolor{darkblue}{rgb}{0.0, 0.0, 0.55}
\usepackage{hyperref}
\hypersetup{
    pdfstartview=Fit,
    linktoc=all,
    colorlinks=true,
    linkcolor=darkblue,
    citecolor=darkblue,
    filecolor=darkblue,
    urlcolor=darkblue,
    bookmarksopen=false,
    bookmarksopenlevel=2,
    bookmarksdepth=3,
} 
\usepackage[all]{hypcap}

\let\sidebysidefigures=\undefined

\fi

\theoremstyle{plain}

\theoremstyle{definition}
\newtheorem{definition}{Definition}

\theoremstyle{remark}

\def\ncalO{{\mathcal{O}}}

\def\ncalS{{\mathcal{S}}}

\def\nbbN{{\mathbb{N}}}

\def\E{\mathbb{E}}
\def\P{\mathbb{P}}
\def\R{\mathbb{R}}

\def\indicator{{\mathbb{1}}}

\def\th{{^\text{th}}}

\def\pl{\mathtt{P_L}}

\def\PPPa{{\Phi}}

\def\pppai{{\lambda}}
\def\pppais{{\tilde{\pppai}}}

\def\N{{\sigma^2}}
\def\threshold{{\beta}}
\def\SINR{{\sf SINR}}

\def\origin{{o}}
\def\pg{{\gamma}}

\def\Arc{\angle}
\def\mArc{\Arc_{\max}}

\renewcommand{\cos}[1]{\rm cos \left ( #1 \right )}
\renewcommand{\sin}[1]{\rm sin \left ( #1 \right )}
\newcommand{\sinsq}[1]{{\rm sin}^2\left ( #1 \right )}
\newcommand{\be}{\begin{equation}}
\newcommand{\ee}{\end{equation}}
\newcommand{\beq}{\begin{eqnarray}}
\newcommand{\eeq}{\end{eqnarray}}

\allowdisplaybreaks
\newlength{\maxfigurewidth}
\newlength{\figurewidth}
\ifx\sidebysidefigures\undefined
\else
\fi

\begin{document}


\title{A Tractable Metric for Evaluating Base Station Geometries in Cellular Network Localization}

\author{%
Javier Schloemann, Harpreet S. Dhillon, and R. Michael Buehrer%
\ifx\arXiv\undefined\else%
	\thanks{This work has been submitted to the IEEE for possible publication. Copyright may be transferred without notice, after which this version may no longer be accessible.}%
\fi%
\thanks{The authors are with the Mobile and Portable Radio Research Group (MPRG), Wireless@Virginia Tech, Blacksburg, VA, USA. Email: \{javier, hdhillon, buehrer\}@vt.edu. 
	\hfill Manuscript last updated: \today.
}%
}


\maketitle

\begin{abstract}

In this letter, we present a new metric for characterizing the geometric conditions encountered in cellular positioning based on the angular spread of the base stations (BSs). The metric is shown to be closely related to the geometric-dilution-of-precision (GDOP), yet has the benefit of being characterizable in terms of the network parameters for BS layouts modeled according to a Poisson point process (PPP). As an additional benefit, the metric is shown to immediately yield a device's probability of being inside or outside the convex hull of the BSs, which localization researchers will widely-recognize as being a strong indicator of localization performance.

\end{abstract}


\section{Introduction}\label{Sec:Introduction}

\IEEEPARstart{W}{ith} the proliferation of smart phones enabling location-based services as well as increased pressure due to federal regulations such as the FCC E911 mandate~\cite{FCCE911CFR}, cellular network localization has recently garnered increasing interest. In order to gain insight into how network parameters affect cellular localization performance, industry has typically relied on simulation-based results. This is because it is a standard practice to model BS deployments using a regular grid model, which does not lend itself well to mathematical analysis. Motivated by the current capacity-driven irregular deployments of low-power base stations (BSs), such as pico and femtocells, there have been several successful attempts at characterizing the downlink and uplink performance of cellular networks by modeling the BS locations according to a homogeneous Poisson point process (PPP) (e.g.,~\cite{Andrews2011, Dhillon2012}). 
In addition to being more realistic in capturing the irregularities of current deployments, the PPP-based models lend tractability by allowing the use of tools from stochastic geometry in network analyses. 

The PPP approach is the basis of a new model proposed in~\cite{Schloemann2015a,Schloemann2015c} for studying cellular positioning. The three fundamental components determining localization performance are (i) the number of participating BSs, (ii) the locations of these BSs relative to the device being localized, and (iii) the accuracy of the positioning observations. In~\cite{Schloemann2015c}, the model was used to obtain accurate analytic expressions of BS \emph{hearability}, i.e. the number of participating BSs. A well-known and fundamental metric for evaluating the geometric component is the \emph{geometric-dilution-of-precision} (GDOP)~\cite{Torrieri1984}. This metric is closely related to localization performance, but is limited to studying \emph{fixed} positioning scenarios. Instead, it is desirable to understand how network parameters affect the favorability of all geometries which may be encountered. This motivates the need for a new tractable metric which can be used to gain insights into how network parameters affect localization geometries.

\section{System Model}\label{P3:Sec:SystemModel}

\subsection{Spatial base station layout}
The locations of the BSs are modeled using a homogeneous PPP $\PPPa \in \R^2$ with density $\pppai$~\cite{Haenggi2013}. Due to the stationarity of a homogeneous PPP, the device to be localized is assumed to be located at the origin $\origin$. For the link from some BS $x \in \PPPa$ to the origin, the received \emph{signal-to-interference-plus-noise ratio} (SINR) is expressed as:
\begin{equation}
\SINR_x = \frac{P \ncalS_{x} \|x\|^{-\alpha}}{\sum_{\substack{y \in \PPPa\\y \neq x}} a_y P \ncalS_{y} \|y\|^{-\alpha} + \sigma^2},
\label{Eq:SINR_k_L}
\end{equation}
where $P$ is the transmit power, $\ncalS_z$ denotes the independent shadowing affecting the signal from BS $z$ to the origin, $\alpha > 2$ is the pathloss exponent, and $\N$ is the noise variance. The BSs are assumed active with probability $f$, which represents the average network load.  To model this loading, $a_y$ are independent indicator random variables modeling the thinning of the interference field by taking on value 1 with probability $f$.  Furthermore, localization systems often include some sort of multiplicative processing gain $\pg$ due to integration, which is assumed to average out the effect of small scale fading.


\subsection{Number of participating base stations}\label{P3:Sec:BSSelection}

For localization, it is well-known that including an increasing number of BSs in the localization procedure results in a general improvement in positioning accuracy. Thus, we assume that for purposes of localization, a device will take advantage of as many BSs as it can successfully detect (or \emph{hear}), i.e., all BSs whose signals arrive with some minimum link quality. Specifically, a BS $x$ participates in the localization procedure when $\pg\cdot\SINR_{x} \geq \threshold$, where $\threshold$ is the post-processing SINR threshold above which the signals from the BSs must arrive in order for them to be successfully detected. Let $\nbbN = \sum_{x \in \PPPa} \indicator(\SINR_x \geq \threshold/\pg)$ represent the number of BSs hearable at the origin. In this work, we employ the hearability results derived in~\cite{Schloemann2015a,Schloemann2015c} so that
\begin{align}
\P(\nbbN = L) = \pl(f,f,\alpha,\threshold,\pg,\pppais) - \mathtt{P_{L+1}}(f,f,\alpha,\threshold,\pg,\pppais),
\end{align}
where $\pppais = \pppai \E[\ncalS_z^{2/\alpha}]$ is the shadowing-transformed density of the original PPP $\PPPa$ and $\pl(f,f,\alpha,\threshold,\pg,\pppais)$, the probability of hearing at least $L$ BSs in interference-limited networks (including cellular networks), is obtained from the hearability expressions presented in~\cite[Theorem~2 and its corollaries]{Schloemann2015c}.


\section{Geometric Characterization}

A particularly nice property of the GDOP is that it is closely related to the \emph{Cr\'{a}mer-Rao lower bound} (CRLB), the lower bound on the mean square error (MSE) of an unbiased position estimator. For TOA-based localization, if the errors in the distance measurements between the mobile and different BSs exhibit a common standard deviation of $\sigma_r$ meters, then the CRLB may be calculated from the GDOP as follows~\cite{Torrieri1984}:
\begin{align}
\text{CRLB} = \sigma_r^2 \cdot \text{GDOP}^2\!\!,\label{P2:Eq:CRLBGDOP}
\end{align}
More specifically, for two-dimensional positioning with position $[x,y]^T$, we can write GDOP as
\begin{equation}
\text{GDOP} = \frac{\sqrt{\sigma_x^2+\sigma_y^2}}{\sigma_r} = \sqrt{tr({\bf H}^T{\bf H})^{-1}}
\end{equation}
\begin{equation}
{\bf H} = \left [ \begin{array}{cc} \frac{x-x_1}{r_1} & \frac{y-y_1}{r_1} \\ \frac{x-x_2}{r_2} & \frac{y-y_2}{r_2}\\ \vdots & \vdots \\ \frac{x-x_L}{r_L} & \frac{y-y_N}{r_L}  \end{array} \right ]
\end{equation}
where $\sigma_x^2$ and $\sigma_y^2$ are the variances of the estimates of $x$ and $y$, $r_i$ is the distance to BS $i$.  While it is not difficult to calculate the GDOP for specific scenarios, we are not aware of any technique for characterizing the {\it distribution} of GDOP over all positioning scenarios which may be encountered for a given set of network parameters.

In order to characterize the geometric conditions of BSs, we now introduce an alternate metric to the GDOP which is based on the angular spread of the BSs from the mobile device.
\begin{definition}[Maximum angular separation metric]
Consider $L$ participating BSs, where the angle from the origin to the $\ell\th$ BS is $\theta_\ell$, measured from some common baseline. For convenience, we number the BSs in order such that $\theta_1 < \theta_2 < \ldots < \theta_L$ and let $\angle_\ell = \theta_{\ell+1}-\theta_\ell$ for $1 \leq \ell < L$ and $\angle_L = 2\pi - (\theta_L - \theta_1)$. 
The metric
\begin{align}
\mArc = \max\{\angle_1, \ldots, \angle_L\}
\end{align}
is then the {\em maximum angular separation} between the BSs.  
\end{definition}
Now, noting that $\frac{x-x_i}{r_i}=\cos{\theta_i}$ and $\frac{y-y_i}{r_i}=\sin{\theta_i}$, we can re-write GDOP as 
\beq
GDOP & = & \frac{\sqrt{L}}{\sqrt{\sum_{i=1}^{L-1}\sum_{j=i+1}^{L}\sinsq{\theta_j-\theta_i}}} \nonumber \\
& = & \frac{\sqrt{L}}{\sqrt{\sum_{i=1}^{L-1}\sum_{j=i+1}^{L}\sinsq{\sum_{k=i}^{j-1}\angle_k}}} \nonumber \\
& \leq & \frac{\sqrt{L}}{|\sin{\angle_{max}}|}
\label{eq:TOA_relation}
\eeq

\begin{figure}
\centering
\includegraphics[width=\figurewidth]{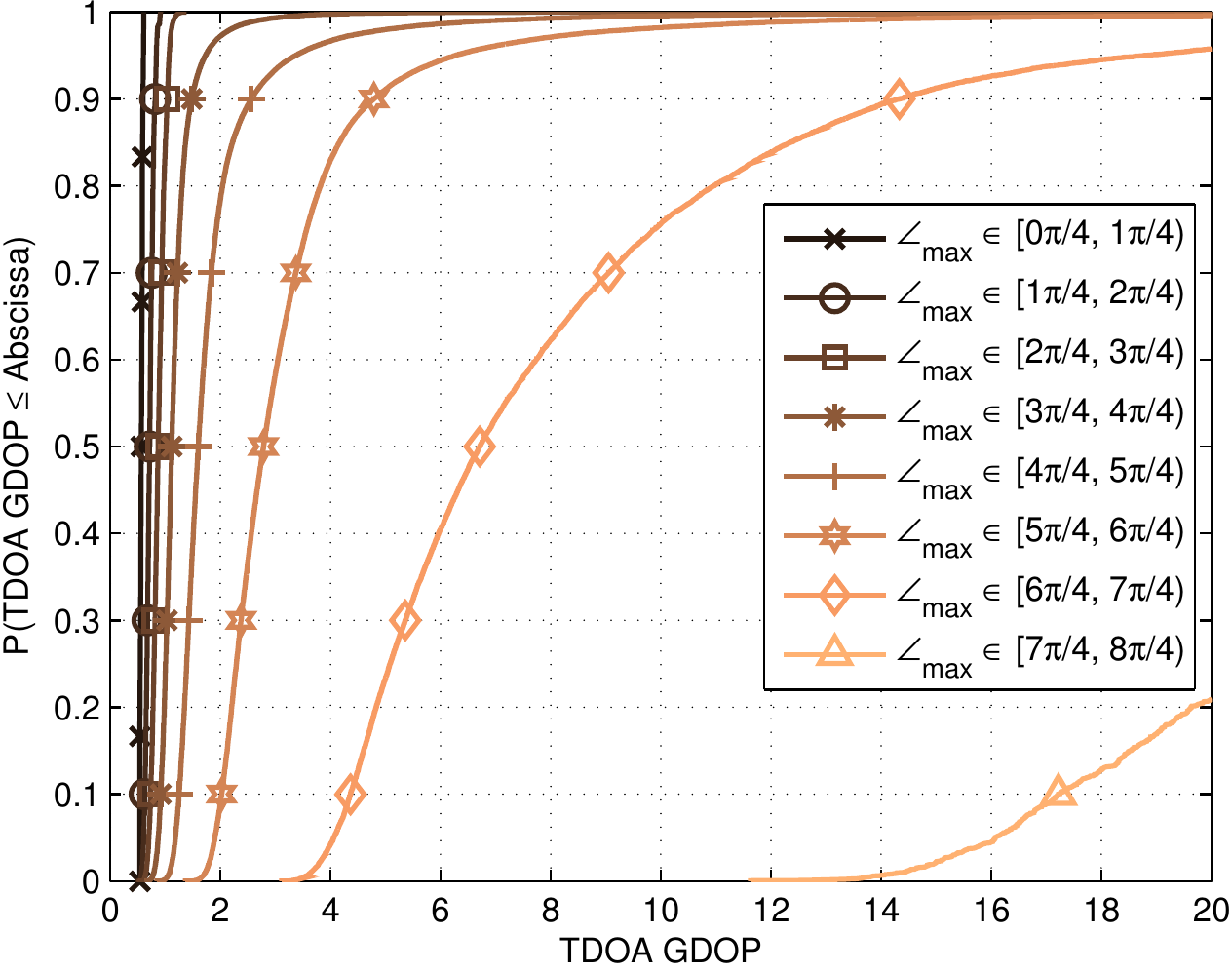}
\caption{\textsc{Relationship between the GDOP and $\mArc$} in cellular TDOA positioning. For small values of $\mArc$, the GDOP is very predictable, while larger values of $\mArc$ provide a good indication of the minimum geometric dilution that can be expected.}
\label{Fig:PPP_FGDOP_v_psi_minL}
\end{figure}

Thus, for TOA we find that there is clearly a relationship between $\mArc$ and GDOP. In this letter, we consider downlink TDOA positioning, e.g., OTDOA in LTE~\cite{Fischer2014}, which is typically the first-choice fallback technique in cellular networks when the Global Positioning System is unavailable. Showing a clean relationship similar to~\eqref{eq:TOA_relation} for TDOA is slightly more complicated than TOA. Thus, in order to compare the distributions of the two metrics for TDOA, we take an alternate correlation-based approach in which we 
simulated 2 million TDOA positioning scenarios with $f=1$, $\alpha=4$, $\threshold/\pg = -10$dB, log-normal shadowing with a standard deviation of $\sigma_s = 8$dB, and a BS density of $\pppai = 2/(\sqrt{3} \cdot 500^2)~\text{m}^{-2}$ which is equivalent to that of an infinite hexagonal grid with 500m intersite distances (ISD).\footnote{Although the analysis is valid for all values of $\alpha>2$, we present results using $\alpha=4$ due to its proximity to the $\alpha=3.76$ value used in 3GPP positioning studies~\cite{R1-091443} and the availability of simplified expressions for hearability in~\cite{Schloemann2015c} for this special case.} In Figure~\ref{Fig:PPP_FGDOP_v_psi_minL}, the distributions of the GDOP are plotted for all scenarios whose $\mArc$ values fell within the specified ranges. Small values of $\mArc$ (approximately $3\pi/4$ and below) are tightly coupled with low GDOP values. As $\mArc$ increases, the actual GDOP values are less predictable, though it is clear that higher $\mArc$ values are tied to higher GDOPs (in fact, the GDOPs can be lower bounded). An alternate way to look at the connection between the metrics is through a correlation analysis. In Table~\ref{Table:P3:correlation}, the Spearman rank correlation~\cite{Spearman1904} and the well-known Pearson's product-moment (linear) correlation between $\mArc$ and the GDOP are shown for several values of $L$. First, we note that the linear correlation between $\mArc$ and GDOP is quite poor. This is not unexpected, considering that it is difficult to draw a linear relationship between one metric whose values are limited to between 0 and $2\pi$ and a second metric which is unbounded. On the other hand, we note that the Spearman rank correlation is quite good (above 0.9 for all values of $L$). This implies several things: (i) that $\mArc$ is a good metric for comparing the geometries of two positioning scenarios and (ii) that the relationship between $\mArc$ and GDOP can be modeled well by some monotonic function. Though determining the best monotonic function which relates $\mArc$ and GDOP is outside the scope of this paper, consider the Pearson correlation between $\mArc$ and $\log(\text{GDOP})$ shown in Table~\ref{Table:P3:correlation}. The fact that there is a significant coupling implies that there exist some values of $m$ and $b$ for which $\text{GDOP} = e^{m\mArc+b}$ provides a simple and reasonable approximate transformation between the two metrics.

\begin{table}
\centering
\caption{Correlation Between $\angle_{\max}$ and GDOP}
\label{Table:P3:correlation}
\begin{tabular}{@{}lcccc@{}}
\toprule
Correlation Metric & $L$ = 4 & $L$ = 5 & $L$ = 6 & $L \geq$ 4 \\
\midrule
Spearman's $\rho$ rank & 0.908 & 0.927 & 0.924 & 0.912 \\
Pearson's, {\sc GDOP} & 0.029 & 0.513 & 0.659 & 0.023 \\
Pearson's, $\log({\textsc{GDOP}})$ & 0.843 & 0.870 & 0.880 & 0.871 \\
\bottomrule
\end{tabular}
\end{table}

\begin{figure}
\centering
\includegraphics[width=\figurewidth]{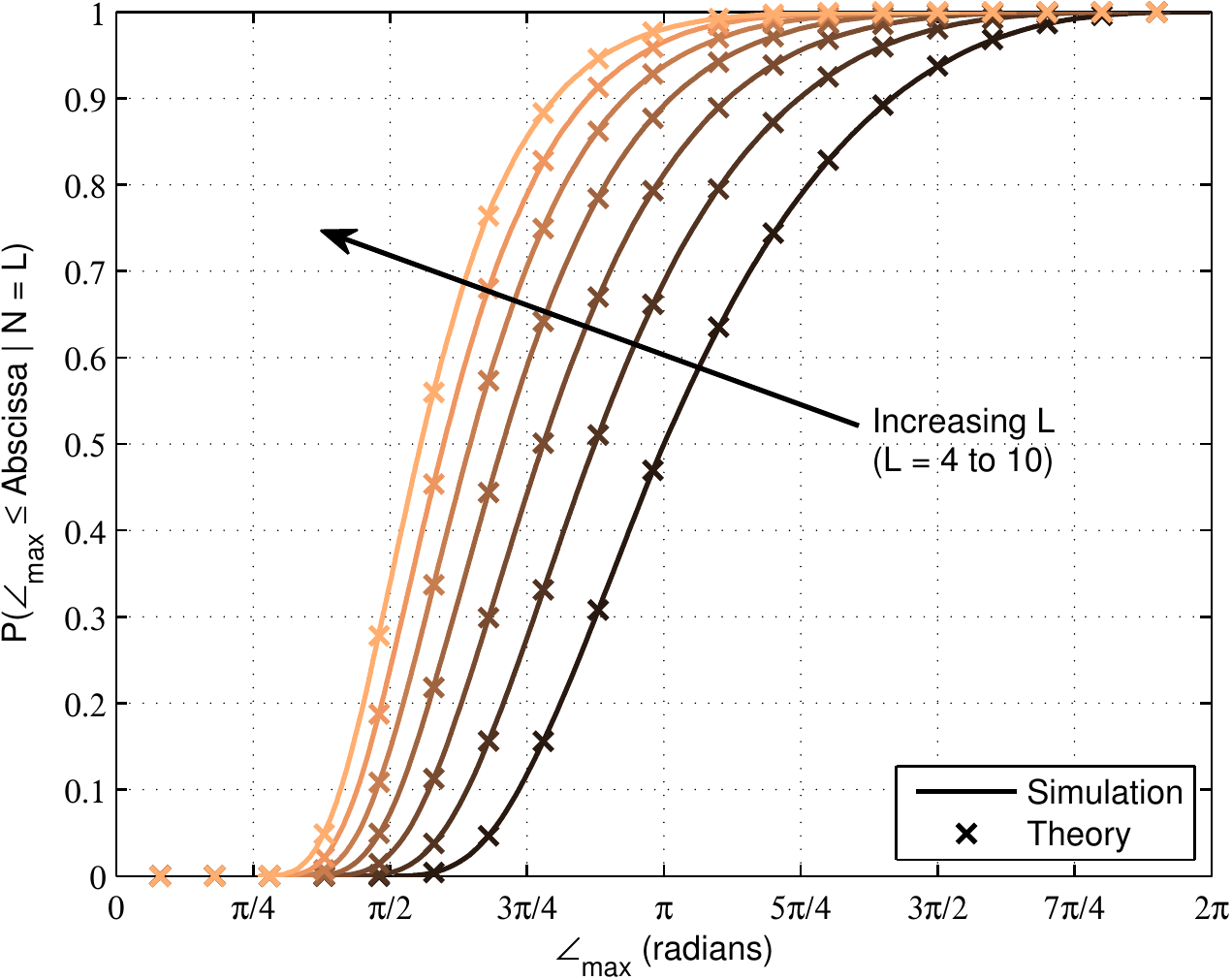}
\caption{\textsc{The impact of the number of BSs}: As more BSs participate in the localization procedure, the distributions of $\mArc$ become increasingly more desirable (i.e., there is a general reduction in the $\mArc$ values encountered).}
\label{Fig:PPP_FPsi_v_L}
\end{figure}

Now that a relationship between the two metrics has been established, we continue our analysis by characterizing the distribution of $\mArc$. What we were seeking is the probability that the maximum angle made by the base station is greater than some value.  Calculating this probability from first principles is not straightforward.  However, (and quite fortunately) this probability is strongly related to two classic problems in probabilistic geometry,  the probability that $L$ random arcs cover the circumference of a circle and the random division of a unit interval (see \cite{Solomon1978} for a discussion).  Concerning the former problem, Stevens derived the probability that $n$ random arcs of size $a$ covered a unit circumference circle in what many consider to be an ingenious proof \cite{Stevens1939}.  It was later shown that this result could be related to the distribution of the number of arcs needed to cover a unit circle and the random division of a unit interval (see \cite{Flatto1962}).  Although space doesn't allow for a full derivation, Stevens' original result can be modified to provide the CDF of $\mArc$.  Specifically, for a fixed value of $L$,
\begin{align}
\P(\mArc \leq \varphi \vert \nbbN = L) = \sum_{n=0}^{\chi} \binom{L}{n} (-1)^n \left(1-\frac{n\varphi}{2\pi}\right)^{L-1}\!\!\!\!\!\!\!\!\!,
\end{align}
where $\chi = \min\left\{L, \left\lfloor 2\pi/\varphi \right\rfloor \right\}$~\cite{Solomon1978}. 
In Figure~\ref{Fig:PPP_FPsi_v_L}, the theoretical distribution of $\P(\mArc \leq \varphi \vert \nbbN = L)$ is plotted for several different values of $L$ and verified to match the $\mArc$ distributions observed in the aforementioned simulation. As expected, higher values of $L$ are desirable for reducing both the minimum and maximum values of $\mArc$. Similarly, higher values of $L$ are known to yield generally lower GDOP values.

Since the distribution of the number of hearable BSs depends on the network parameters, it is worthwhile to characterize the distribution of $\mArc$ across different $L$ values weighted by the probability of hearing $L$ BSs. Consider only those scenarios where $\nbbN \geq 4$, i.e., those times when sufficient BSs are hearable to ensure unique localizability using TDOA~\cite{Buehrer12}. Then,
\begin{align}
\P(\mArc \leq \varphi &\vert \nbbN \geq 4)
= \frac{\P(\mArc \leq \varphi, \nbbN \geq 4)}{\P(\nbbN \geq 4)} \notag\\
&=\!\sum_{L=4}^\infty \frac{\P(\mArc \leq \varphi, \nbbN = L)}{\P(\nbbN \geq 4)} \notag\\
&=\!\sum_{L=4}^\infty \frac{\P(\mArc \leq \varphi \vert \nbbN = L)\P(\nbbN = L)}{\P(\nbbN \geq 4)}.\label{Eq:P3:FmArc}
\end{align}
Note that although the summation limits are infinite, it is not difficult to calculate \eqref{Eq:P3:FmArc} with a high accuracy since $\P(\nbbN = L) \to 0$ for even relatively low values of $L$ (e.g., see Figure~11 in~\cite{Schloemann2015c}). While this behavior eases computation, it makes it increasingly more difficult to obtain lower values of $\mArc$ as the expected number of BSs required to satisfy $\mArc \leq \varphi$ is
\begin{align}
	\E[L] = \sum_{n=1}^{\lfloor \varphi/2\pi - 1 \rfloor} (-1)^n \cdot \frac{\left(1 - \frac{n\varphi}{2\pi}\right)^{n-1}}{\left(\frac{n\varphi}{2\pi}\right)^{n+1}},\notag
\end{align}
which grows on the order $\ncalO\left(\varphi^{-1} \log \varphi^{-1}\right)$ as $\varphi \to 0$~\cite{Flatto1962}.\footnote{The term $\ncalO$ denotes the standard big-O notation which is used to describe the asymptotic behavior of a function.}

\section{Numerical Results and Discussion}\label{Sec:NumericalResultsAndDiscussion}

\begin{figure}
\centering
\includegraphics[width=\figurewidth]{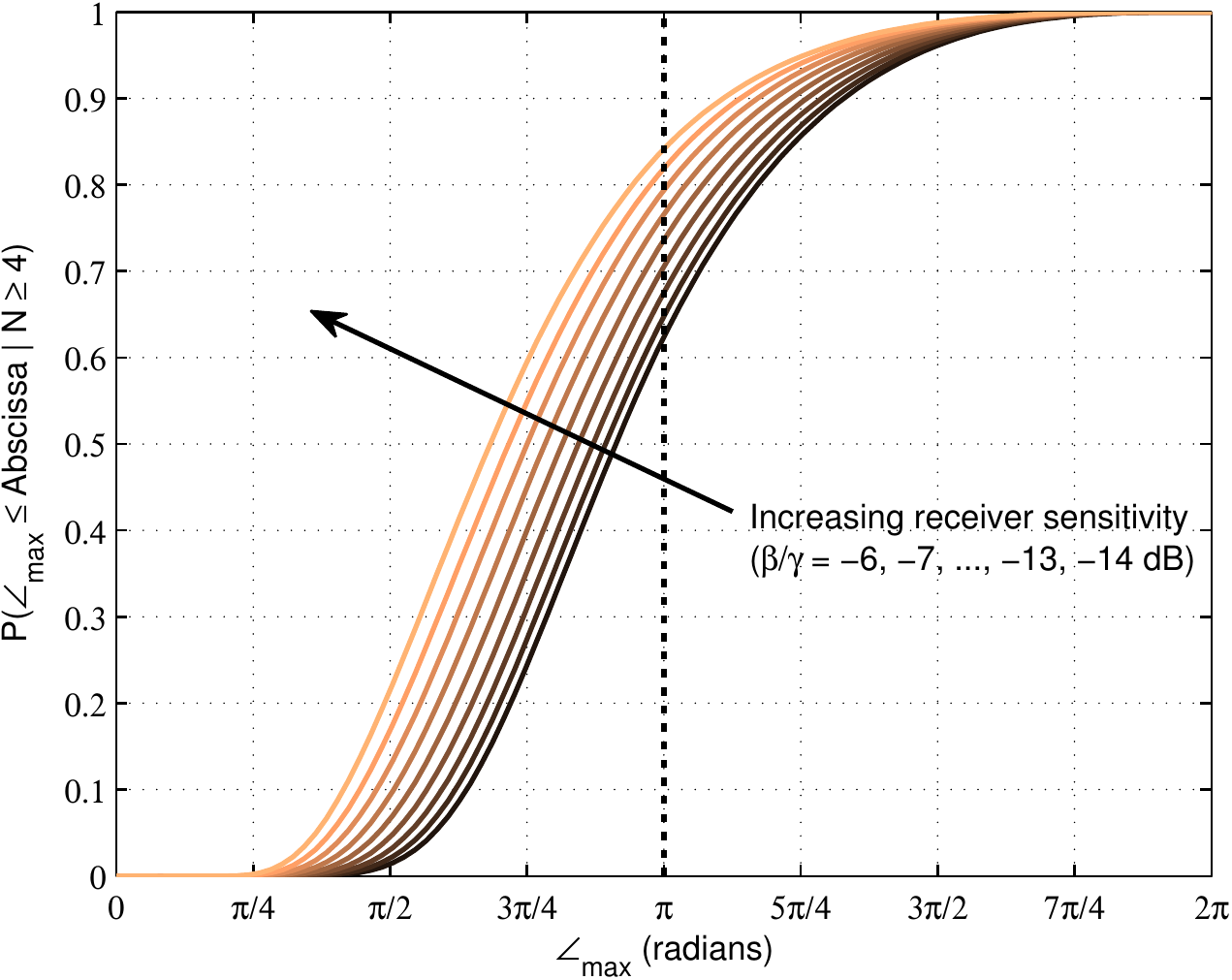}
\caption{\textsc{The impact of receiver sensitivity}: While being able to detect BS signals at lower SINRs generally improves hearability, the effect on geometry is not as pronounced as might be expected ($f=0.5, \alpha=4, \sigma_s=8\text{dB}, \pppai = 2/(\sqrt{3} \cdot 500^2)~\text{m}^{-2}$).}
\label{Fig:FmaxAngle_v_targetSINR}
\end{figure}

\begin{figure}
\centering
\includegraphics[width=\figurewidth]{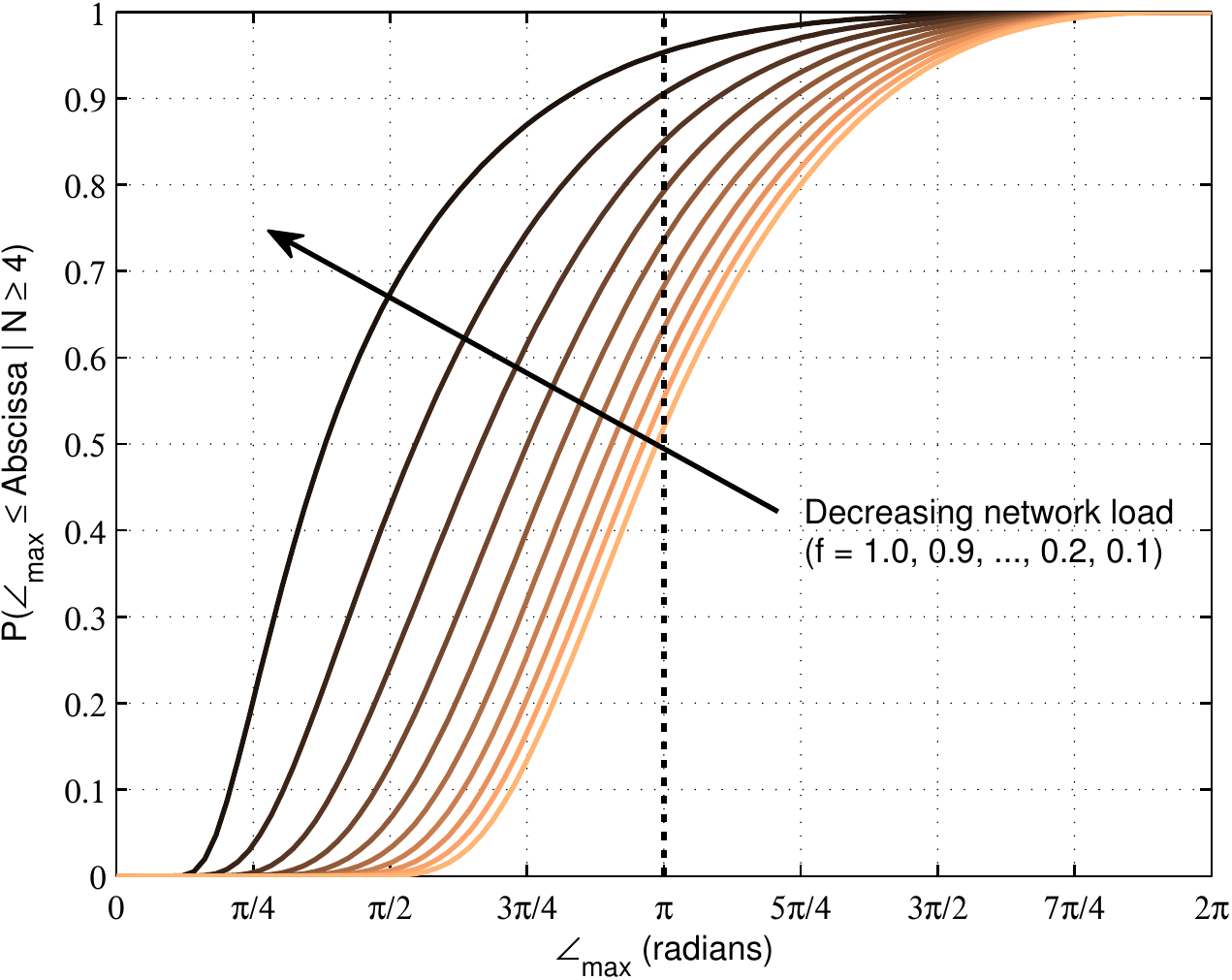}
\caption{\textsc{The impact of average network load}: Reducing the network load yields a significant improvement of the distribution of $\mArc$. In particular, by considering $\mArc \leq \pi$, we observe a drastic improvement in the probability of being inside the convex hull of the BSs ($\threshold/\pg=-10\text{dB}, \alpha=4, \sigma_s=8\text{dB}, \pppai = 2/(\sqrt{3} \cdot 500^2)~\text{m}^{-2}$).}
\label{Fig:FmaxAngle_v_q}
\end{figure}

\begin{figure}
\centering
\includegraphics[width=\figurewidth]{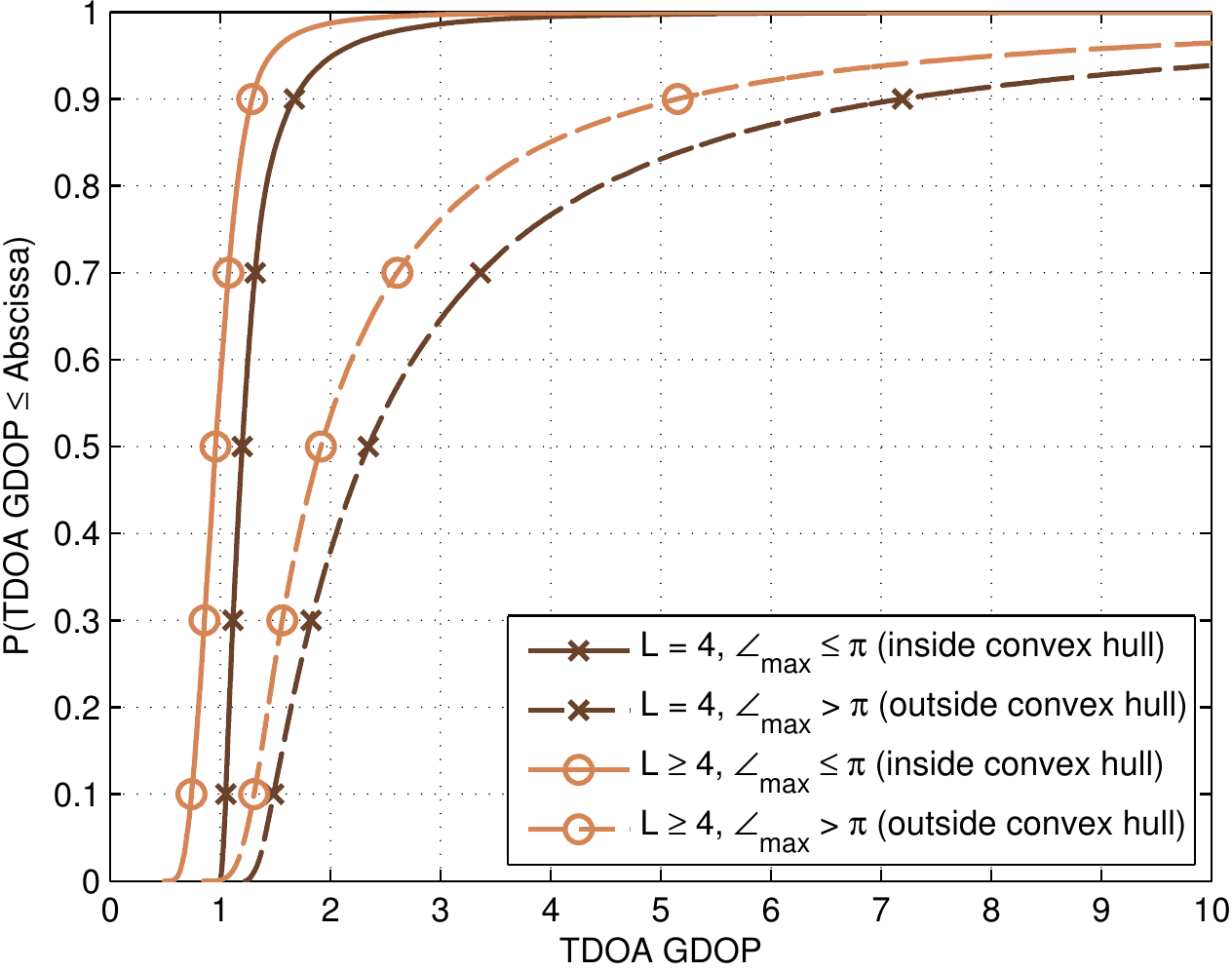}
\caption{\textsc{Localization performance inside and outside the convex hull}: This figure illustrates the value of being inside the convex hull, where GDOP values are almost certainly below 4 ($\threshold/\pg=-10\text{dB}, \alpha=4, \sigma_s=8\text{dB}, \pppai = 2/(\sqrt{3} \cdot 500^2)~\text{m}^{-2}$).}
\label{Fig:PPP_FGDOP_v_convexhull}
\end{figure}

In this section, we present some insights into how the network parameters affect the distribution of $\mArc$ and also reveal a particularly valuable geometric insight which is easily obtained from the $\mArc$ metric. First, let us consider how altering the required pre-processing SINR, $\threshold/\pg$, which is a measure of detection sensitivity, affects the distribution of $\mArc$. In Figure~\ref{Fig:FmaxAngle_v_targetSINR}, \eqref{Eq:P3:FmArc} is plotted for target pre-processing SINRs between $-6$dB and $-14$dB (the range of values suggested at different points throughout 3GPP positioning studies). While hearability clearly improves as shown in~\cite{Schloemann2015c}, the effect on the distribution of $\mArc$ is surprisingly small. In Figure~\ref{Fig:FmaxAngle_v_q}, we again consider \eqref{Eq:P3:FmArc}, but now versus changes in the average network load $f$. We see that reducing the network load has a significantly greater effect on the geometric conditions measured using $\mArc$. Consider the values at $\P(\mArc \leq \pi|\nbbN \geq 4)$ highlighted by the dashed lines in Figures~\ref{Fig:FmaxAngle_v_targetSINR} and~\ref{Fig:FmaxAngle_v_q}. As the network load decreases from $f=1$ to $f=0.1$, $\P(\mArc \leq \pi|\nbbN \geq 4)$ increases from 0.5 to 0.95. This is significant because by evaluating $\P(\mArc \leq \pi|\nbbN \geq 4)$, one immediately obtains the \emph{probability of being inside the convex hull of the BSs}, which is known to be a strong indicator of localization performance~\cite{Ash2008,Savvides2003}. This is demonstrated in Figure~\ref{Fig:PPP_FGDOP_v_convexhull} for both $L=4$ and $L\geq 4$, which reveals a drastic increase in the GDOPs when the device to be localized is outside the convex hull of the BSs compared to when it is inside the convex hull, where the GDOP values are essentially upper-bounded by 4.

\section{Conclusion}
In this letter, we presented an alternate metric to the well-known GDOP for characterizing the geometric conditions of the base stations in cellular localization. The metric is based on the maximum angular separation between BSs, which was shown to be closely related with the GDOP. In contrast to the GDOP, which is limited to studying the geometric conditions of \emph{fixed} positioning scenarios, our metric can be tractably characterized in terms of the network parameters across \emph{all} positioning scenarios. Additionally, the metric can be employed to immediately obtain the probability that a device will be inside or outside the convex hull of the BSs, which is a widely-recognized indicator of localization performance.

\bibliographystyle{IEEEtran}
\bibliography{paper_3}

\begin{thebibliography}{10}
\providecommand{\url}[1]{#1}
\csname url@samestyle\endcsname
\providecommand{\newblock}{\relax}
\providecommand{\bibinfo}[2]{#2}
\providecommand{\BIBentrySTDinterwordspacing}{\spaceskip=0pt\relax}
\providecommand{\BIBentryALTinterwordstretchfactor}{4}
\providecommand{\BIBentryALTinterwordspacing}{\spaceskip=\fontdimen2\font plus
\BIBentryALTinterwordstretchfactor\fontdimen3\font minus
  \fontdimen4\font\relax}
\providecommand{\BIBforeignlanguage}[2]{{%
\expandafter\ifx\csname l@#1\endcsname\relax
\typeout{** WARNING: IEEEtran.bst: No hyphenation pattern has been}%
\typeout{** loaded for the language `#1'. Using the pattern for}%
\typeout{** the default language instead.}%
\else
\language=\csname l@#1\endcsname
\fi
#2}}
\providecommand{\BIBdecl}{\relax}
\BIBdecl

\bibitem{FCCE911CFR}
{Code of Federal Regulations}, ``{911 Service},'' 47 C.F.R. 20.18(h)(2)(ii),
  2015.

\bibitem{Andrews2011}
J.~G. Andrews, F.~Baccelli, and R.~K. Ganti, ``{A tractable approach to
  coverage and rate in cellular networks},'' \emph{IEEE Trans. Commun.},
  vol.~59, no.~11, pp. 3122--3134, Nov. 2011.

\bibitem{Dhillon2012}
H.~S. Dhillon, R.~K. Ganti, F.~Baccelli, and J.~G. Andrews, ``{Modeling and
  analysis of K-tier downlink heterogeneous cellular networks},'' \emph{IEEE J.
  Sel. Areas Commun.}, vol.~30, no.~3, pp. 550--560, Apr. 2012.

\bibitem{Schloemann2015a}
J.~Schloemann, H.~S. Dhillon, and R.~M. Buehrer, ``{Localization performance in
  cellular networks},'' in \emph{Proc. IEEE Int. Conf. Commun. Work. Adv. Netw.
  Localization Navig.}, London, UK, Jun. 2015.

\bibitem{Schloemann2015c}
------, ``{Towards a tractable analysis of localization fundamentals in
  cellular networks},'' \emph{accepted for publication in IEEE Transactions on
  Wireless Communications}, \emph{arXiv:1502.06899 [cs.IT]}.

\bibitem{Torrieri1984}
D.~J. Torrieri, ``{Statistical theory of passive location systems},''
  \emph{IEEE Trans. Aerosp. Electron. Syst.}, vol. AES-20, no.~2, pp. 183--198,
  Mar. 1984.

\bibitem{Haenggi2013}
M.~Haenggi, \emph{{Stochastic Geometry for Wireless Networks}}.\hskip 1em plus
  0.5em minus 0.4em\relax New York: Cambridge University Press, 2013.

\bibitem{Fischer2014}
S.~Fischer, ``{Observed Time Difference Of Arrival (OTDOA) positioning in 3GPP
  LTE},'' \emph{Qualcomm White Pap.}, 2014.

\bibitem{R1-091443}
{Third Generation Partnership Project (3GPP)}, ``{R1-091443: Evaluation
  parameters for positioning studies},'' Alcatel-Lucent, Ericsson, Motorola,
  Nokia, NSN, Nortel, Qualcomm Europe, 3GPP TSG-RAN WG1 \#56bis, Seoul, Korea,
  Mar. 2009.

\bibitem{Spearman1904}
C.~Spearman, ``{The proof and measurement of association between two things},''
  \emph{Am. J. Psychol.}, vol.~15, no.~1, pp. 72--101, 1904.

\bibitem{Solomon1978}
H.~Solomon, \emph{{Geometric Probability}}.\hskip 1em plus 0.5em minus
  0.4em\relax Society for Industrial and Applied Mathematics, 1978.

\bibitem{Stevens1939}
W.~L. Stevens, ``{Solution to a geometrical problem in probability},''
  \emph{Ann. Eugen.}, vol.~2, pp. 315--320, 1939.

\bibitem{Flatto1962}
L.~Flatto and A.~G. Konheim, ``{The random division of an interval and covering
  of a circle},'' \emph{SIAM Rev.}, vol.~4, no.~3, pp. 211--222, 1962.

\bibitem{Buehrer12}
R.~M. Buehrer and S.~Venkatesh, ``{Fundamentals of time-of-arrival-based
  position location},'' in \emph{Handbook of Position Location: Theory,
  Practice, and Advances}, S.~A. Zekavat and R.~M. Buehrer, Eds.\hskip 1em plus
  0.5em minus 0.4em\relax Hoboken, NJ: IEEE Press/John Wiley and Sons, 2012.

\bibitem{Ash2008}
J.~N. Ash and R.~L. Moses, ``{On optimal anchor node placement in sensor
  localization by optimization of subspace principal angles},'' \emph{ICASSP,
  IEEE International Conference on Acoustics, Speech and Signal Processing -
  Proceedings}, pp. 2289--2292, 2008.

\bibitem{Savvides2003}
A.~Savvides, W.~Garber, S.~Adlakha, R.~Moses, and M.~B. Srivastava, ``{On the
  error characteristics of multihop node localization in ad-hoc sensor
  networks},'' in \emph{Proc. 2nd Int. Conf. Inf. Process. Sens. Networks},
  Apr. 2003, pp. 317--332.

\end{thebibliography}

\end{document}